\documentclass[aps,prl,twocolumn]{revtex4}

\usepackage{amsmath}
\usepackage{amssymb}
\usepackage{graphicx}
\def\sigmaxc{\stackrel{\leftrightarrow}{\sigma}}

\begin{document}

\title{Stochastic Time-Dependent Current-Density-Functional Theory}

\author{Massimiliano Di Ventra}
\email{diventra@physics.ucsd.edu}
\affiliation{University of California - San Diego, La Jolla,
California 92093}

\author{Roberto D'Agosta}
\email{dagosta@physics.ucsd.edu}
\affiliation{University of
California - San Diego, La Jolla, California 92093}

\date{\today}

\begin{abstract}
A time-dependent current-density-functional theory for many-particle
systems in interaction with arbitrary external baths is developed.
We prove that, given the initial quantum state $|\Psi_0\rangle$ and
the particle-bath interaction operator, two external vector
potentials ${\bf A}({\bf r},t)$ and ${\bf A}'({\bf r},t)$ that
produce the same ensemble-averaged current density, $\overline{ {\bf
j}({\bf r},t)}$, must necessarily coincide up to a gauge
transformation. This result greatly expands the applicability of
time-dependent density-functional theory to open quantum systems,
and allows for first-principles calculations of many-particle time
evolution beyond Hamiltonian dynamics.
\end{abstract}

\maketitle

Time-dependent density-functional theory (TDDFT)~\cite{Runge1984} is
becoming an important tool in the study of the dynamics of
many-particle systems~\footnote{See, e.g., M.A.L. Marques, C.A.
Ullrich, F. Nogueira, A. Rubio, K. Burke, and E. K. U. Gross, in
{\it Time-Dependent Density Functional Theory}, vol {\bf 706} of
{\it Lecture Notes in Physics} (Springer, Berlin, 2006), for an
up-to-date state-of-the-art of the theory and its applications.}. The
theory, which extends the applicability of ground-state
density-functional theory~\cite{Hohenberg1964,Kohn1965} to
non-equilibrium problems, rests solidly on a theorem proved by Runge
and Gross (RG)~\cite{Runge1984}. The RG theorem shows that, for a
fixed initial quantum state $|\Psi_0\rangle$, the external
time-dependent scalar potential $V_{ext}({\bf r},t)$ acting on the
many-particle system is uniquely determined (up to a constant) by
the time-dependent single particle density $n({\bf r},t)$. It was
subsequently realized by Vignale and Kohn~\cite{Vignale1996b} that,
since the exchange-correlation (xc) scalar potential in TDDFT is a
strongly non-local functional of the density, a theory formulated in
terms of the current density ${\bf j}({\bf r},t)$ allows for a local
gradient expansion of the xc {\em vector}
potential ${\bf A}_{xc}({\bf r},t)$. This theory goes under the name of
time-dependent current density-functional theory
(TDCDFT)~\cite{Ghosh1988,Vignale1996b}, and has been applied to a variety of
contexts ranging from optical spectra of solids \cite{Berger2005}
and atoms~\cite{Ullrich2001}, to dielectric properties of
polymers~\cite{vanFaassen2002}, to transport in nanoscale
junctions~\cite{Sai2005,DAgosta2006a,DAgosta2006c,Sai2007}, to the
study of systems with memory and/or dissipation
~\cite{Wijewardane2005, DAgosta2006, Kurzweil2004}.

Despite the enormous success that the above time-dependent
density-functional based methods have been experiencing, one needs
to acknowledge that they can strictly deal with particle systems
evolving under Hamiltonian dynamics. There is, however, a large
class of physical problems where one needs to consider interactions
with an external bath. Examples include dephasing of a quantum
system coupled to a boson field, spontaneous emission due to
coupling with the zero-point energy fluctuations, non-radiative
decay, etc. A TDCDFT able to deal with such open-system problems,
without reverting to consider the microscopic dynamics of the bath,
would be a tremendous asset in the study of many-particle problems
beyond Hamiltonian dynamics~\footnote{A first step in this direction
was taken in K. Burke, R. Car, and R. Gebauer, Phys. Rev. Lett. {\bf
94}, 146803 (2005).  Here TDDFT is used in combination with a master
equation to study electronic transport in the presence of
dissipation. However, due to the dependence of the Kohn-Sham
Hamiltonian $\hat H_{KS}$ on the density, a Kohn-Sham master
equation can not be easily derived, because for any operator $\hat
O$ one has in general $\overline {[\hat H_{KS},\hat O]}\neq [\hat
H_{KS},\overline {\hat O}]$.}.

In this Letter, we introduce such a theory which we name
``Stochastic TDCDFT'' and prove a fundamental existence theorem
comparable to the theorems of TDDFT and TDCDFT. Our theorem allows
us to rigourously map the many-particle dynamics in the presence of
an external bath into an effective single-particle dynamics in the
presence of the same bath.

In non-equilibrium quantum statistical mechanics,
the effect of a bath is effectively described using stochastic processes: the
fluctuations introduced by the presence of the bath are accompanied by
dissipative effects in the particle dynamics. Our starting point is
therefore the {\it stochastic} time-dependent Schr\"{o}dinger
equation, which reads
($\hbar=1$)~\cite{vanKampen}
\begin{equation}
\partial_t \Psi({\bf r},t)=-i\hat H(t)\Psi({\bf r},t)-\frac12 \hat V^\dagger \hat V
\Psi({\bf r},t)+\hat V\ell(t)\Psi({\bf r},t)
\label{evolution}
\end{equation}
where $\ell$ describes a stochastic process, and $\hat V$ is an
operator which describes the bath, and its interaction with the
many-particle system~\footnote{The operator $\hat V$ can always be
chosen traceless. Moreover, we can generalize the theory
to more than one bath, each described by an operator ${\hat
V}_{\alpha}$, by solving
$\partial_t \Psi({\bf r},t)=\left[-i \hat H-\frac12
\sum_{\alpha}{\hat V}_{\alpha}^\dagger {\hat V}_{\alpha}\Psi({\bf
r},t) +\sum_{\alpha}{\hat V}_{\alpha}\ell_{\alpha}(t)\right]\Psi({\bf
r},t),$
with $\overline{\ell_{\alpha}(t)}=0$ and
$\overline{\ell_{\alpha}(t)\ell_{\beta}(t')}=\delta_{\alpha
\beta}\delta(t-t')$.}.
In the following we will assume that $\hat V$ is independent
of time and uniform in space.
These restrictions can be easily removed (provided one adds the requirement
that $\hat V$ admits a series expansion in time) and our theorem remains
valid, but we maintain them for the sake of simplicity.
A few examples of the
operator $\hat V$ can be found in~\cite{vanKampen}.
The Hamiltonian operator is
\begin{equation}
\hat H(t)=\sum_{i}\frac{\left[\hat p_i+e
    {\bf A}(\hat r_i,t)\right]^2}{2m}+\frac12 \sum_{i\not = j}U\left(\hat r_i-\hat
r_j\right), \label{h}
\end{equation}
with ${\bf A}(\hat r_i,t)$ a vector potential and $U\left(\hat
r_i-\hat r_j\right)$ a potential describing two-particle
interactions~\footnote{We are working in a gauge in which the
dynamical external scalar potential is zero at any time.}. Without
loss of generality the stochastic process, $\ell(t)$, is chosen such
that it has both zero ensemble average and $\delta-$autocorrelation,
i.e.,
\begin{equation}
\overline{\ell(t)}=0;\;
\overline{\ell(t)\ell(t')}=\delta(t-t'),\label{av}
\end{equation}
where the symbol $\overline{\cdots}$ indicates the average over a
statistical ensemble of identical systems all prepared in the {\it
same} initial quantum state $|\Psi_0\rangle$
\footnote{The theorem may not necessarily hold for stochastic processes
with arbitrary higher moments.}.

The last term on the rhs of Eq.~(\ref{evolution}) describes
precisely the ``fluctuation'' induced by the bath; the second term
is the compensating ``dissipative'' part. Equation~(\ref{evolution}), with the stochastic
process defined in~(\ref{av}),
is a stochastic differential equation which preserves the
ensemble-averaged norm and the
orthogonality between any two states, $\psi_{\alpha}({\bf r},t)$ and $\psi_{\beta}({\bf r},t)$,
$
\int d{\bf r}~\overline{\psi_{\alpha}^* ({\bf r},t)\psi_{\beta}({\bf
r},t)}=\delta_{\alpha \beta}.
\label{ortho}
$
Ensemble-averaged orthonormality can be easily demonstrated by
integrating Eq.~(\ref{evolution}) in a small interval of time
$\Delta t$, taking the scalar product between two
states, and ensemble-averaging this product using the
properties~(\ref{av})~\cite{vanKampen}.

Given any operator in the Heisenberg representation,
its average over the initial state is obtained via
the standard quantum mechanical definition
$\langle \hat O(t)\rangle\equiv\langle \Psi_0|\hat O(t)|\Psi_0\rangle.$
If $\hat O$ is the density operator $\hat n$, one can easily show that the dynamics induced
by Eq.~(\ref{evolution}) preserves the ensemble-averaged particle
number, i.e., the quantity $\overline{N}\equiv\int d{\bf r}~\langle \overline{{\hat
n}({\bf r},t)}\rangle\equiv\int d{\bf r}~\langle \overline{\psi^\dagger
({\bf r},t)\psi({\bf r},t)}\rangle$ is a constant of motion
\cite{vanKampen}. Physically, this corresponds to the case in 
which the bath may exchange particles with the system, while
keeping their average number constant.
Similarly, one can prove that, given any observable $\hat O(t)$, its ensemble
average time evolution, $\overline{\hat O(t)}$, satisfies~\cite{vanKampen}
\begin{equation}
\partial_t \overline{\hat O}=i\left[\hat H,\overline{\hat O}\right]-\frac12
\hat V^\dagger \hat V \overline{\hat O}-  \frac12 \overline{\hat O}
\hat V^\dagger \hat V +\hat V^\dagger \overline{\hat O} \hat V.
\label{operator-dynamics}
\end{equation}
In passing, we note that if $[\overline{\hat O},\hat V]\equiv 0$ at
any time, then the equation of motion (\ref{operator-dynamics})
reduces to $\partial_t \overline{\hat O}=i\left[\hat
H,\overline{\hat O}\right]$, i.e., the presence of the bath does not
affect, on average, the evolution of the observable ${\hat O}$.
Since, in general, $[\hat H,\hat V]\neq 0$, the total energy is not
conserved. Additionally, if ${\hat \rho}$ is the density matrix
operator, from~(\ref{evolution}) we get the well-known (Lindblad)
quantum master  equation
\begin{equation}
\partial_t {\hat \rho}=-i\left[\hat H,{\hat \rho}\right]-\frac12
\hat V^\dagger \hat V{\hat \rho}-  \frac12 {\hat \rho} \hat
V^\dagger \hat V +\hat V {\hat \rho} \hat V^\dagger.
\label{MLouville}
\end{equation}

We define the ensemble-average density
\begin{equation}
\overline {n({\bf r},t)}=\langle \overline{{\hat n}({\bf
r},t)}\rangle,
\end{equation}
and current density
\begin{equation}
\overline {{\bf j}({\bf
r},t)}=\langle \overline{\hat j({\bf r},t)}\rangle,
\end{equation}
where the current operator is defined as
$\hat j({\bf r},t)=\frac{1}{2}\sum_i\left\{\delta({\bf r}-\hat r_i), \hat v_i\right\}$
with
\begin{equation}
\hat v_i=\frac{\hat p_i + e {\bf A}(\hat r_i,t)}m,
\label{velocity}
\end{equation}
the velocity operator of particle $i$, and the symbol $\{\hat A,
\hat B\}\equiv (\hat A \hat B + \hat B \hat A)$ is the
anticommutator of any two operators $\hat A$ and $\hat B$.

We have now set the stage to
formulate and demonstrate the following theorem.

{\it Theorem:} Consider a many-particle system described by the
dynamics in Eq.~(\ref{evolution}) with the many-body Hamiltonian
given by Eq.~(\ref{h}). Let $\overline {n({\bf r},t)}$ and
$\overline {{\bf j}({\bf r},t)}$ be the ensemble-averaged
single-particle density and current density, respectively, with
dynamics determined by the external vector potential ${\bf A}({\bf
r},t)$ and bath $\hat V$. Under reasonable physical assumptions,
given an initial condition $|\Psi_0\rangle$, and an assigned bath operator $\hat V$,
another external potential  ${\bf A}'({\bf r},t)$ which gives the
same ensemble-averaged current density, must necessarily coincide,
up to a gauge transformation, with ${\bf A}({\bf r},t)$.

{\it Proof:} To prove this statement we follow a line of reasoning
commonly used to establish similar theorems in TDDFT and TDCDFT~\cite{vanLeeuwen1999,Vignale2004}. Let us assume that the same
ensemble-averaged density $\overline {n({\bf r},t)}$ and current
density $\overline {{\bf j}({\bf r},t)}$ are also obtained from
another many-particle system with Hamiltonian
\begin{equation}
\hat H'(t)=\sum_{i} \frac{\left[\hat p_i+e
    {\bf A}'(\hat r_i,t)\right]^2}{2m}+\frac12 \sum_{i\not = j}U'\left(\hat r_i-\hat
r_j\right), \label{h1}
\end{equation}
evolving from an initial state $|\Psi'_0\rangle$ and following the
stochastic Schr\"{o}dinger equation~(\ref{evolution}) with the {\em
same} bath operator $\hat V$. $|\Psi'_0\rangle$ gives, in the primed system, the
same initial current and particle densities as in the unprimed system.

The core of the demonstration is as follows: By writing the
equations of motion for $\overline {{\bf j}({\bf r},t)}$ determined
by both ${\bf A}({\bf r},t)$ and ${\bf A}'({\bf r},t)$, we obtain an
equation of motion for the potential difference $\Delta {\bf A}({\bf
r},t)={\bf A}({\bf r},t) - {\bf A}'({\bf r},t)$. We then prove that
$\Delta {\bf A}({\bf r},t)$ is completely determined by the initial
condition via a series expansion in time about $t=0$. Finally, if
the two systems coincide then the unique solution is $\Delta {\bf
A}\equiv 0$, up to a gauge transformation.

The equation of motion for the ensemble-averaged current density is
easily obtained from the equation of motion of the
current density operator. From Eq. (\ref{operator-dynamics}) we get
\begin{eqnarray}
\partial_t \overline {{\bf j}({\bf
r},t)}&=&\frac{\overline {n({\bf r},t)}}{m} \partial_t {\bf A}({\bf
r},t) -\frac{\overline {{\bf j}({\bf
r},t)} }{m}\times\left[\nabla \times {\bf A}({\bf r},t)\right]\nonumber\\
&&\label{currentq}+\frac{\langle \overline{\hat {\mathcal F}({\bf
r},t)}\rangle}{m} +\langle\overline{\hat {\mathcal G}({\bf
r},t)}\rangle\label{currenteq}
\end{eqnarray}
where we have defined
\begin{equation} \label{forces}
\begin{split}
\hat {\mathcal G}({\bf r},t)&=\hat V^\dagger \hat j({\bf r},t) \hat
V  -\frac12 \hat j({\bf r},t) \hat V^\dagger \hat V
-\frac12 \hat V^\dagger \hat V \hat j({\bf r},t),\\
\hat {\mathcal F}({\bf r},t)&=-\sum_{i\not = j}\delta({\bf r}-\hat r_i)
 \nabla_j U\left(\hat r_i-\hat r_j\right)+m \nabla
\cdot \hat \sigmaxc({\bf r},t)
\end{split}
\end{equation}
with the stress tensor $\hat \sigmaxc({\bf r},t)$ given by
\begin{equation}
\hat \sigma_{i,j}({\bf r},t)=-\frac{1}{4}\sum_k \{\hat v_i,\{\hat
v_j,\delta({\bf r}-\hat r_k)\}\}.
\label{stresstensor}
\end{equation}
The first two terms on the rhs of Eq.~(\ref{currenteq}) describe the
effect of the applied electromagnetic field on the dynamics of the
many-particle system; the third is due to particle-particle
interactions while the last one is the ``force'' density exerted by
the bath on the system.

Equations similar to Eqs.~(\ref{velocity}) -- (\ref{stresstensor})
can now be written for the system with the vector potential ${\bf
A}'({\bf r},t)$. Similar force terms $\mathcal{F}'$ and
$\mathcal{G}'$ appear in these new equations. $\mathcal{F}'$ and
$\mathcal{G}'$ differ from the same forces in the unprimed system,
since the initial state, the external vector potentials and the
velocity $\hat v$ are different. By assumption, the
ensemble-averaged current and particle densities are the same in the
two systems, thus
\begin{eqnarray}
\partial_t \overline {{\bf j}({\bf
r},t)}&=&\frac{\overline {n({\bf r},t)}}{m} \partial_t {\bf A}'({\bf
r},t) -\frac{\overline {{\bf j}({\bf
r},t)} }{m}\times\left[\nabla \times {\bf A}'({\bf r},t)\right]\nonumber\\
&&\label{currentq1}+\frac{\langle \overline{\hat {\mathcal F}'({\bf
r},t)}\rangle}{m} +\langle\overline{\hat {\mathcal G}'({\bf
r},t)}\rangle\label{currentqp}.
\end{eqnarray}
Taking the difference of Eqs.~(\ref{currenteq}) and
(\ref{currentqp}) we arrive at
\begin{equation}
\overline {n({\bf r},t)} \partial_t \Delta A(r,t) =\overline {{\bf
j}({\bf r},t)} \times\left[\nabla \times \Delta {\bf A}({\bf
r},t)\right]+\Delta Q({\bf r},t) \label{difference}
\end{equation}
where $\Delta {\bf A}({\bf r},t)\equiv {\bf A}'({\bf r},t)-{\bf
A}({\bf r},t)$ and $\Delta Q({\bf r},t)\equiv Q'({\bf r},t)-Q({\bf
r},t)$ with
$Q({\bf r},t)=\langle \overline{\hat {\mathcal F}({\bf
r},t)}\rangle +m\langle\overline{\hat {\mathcal G}({\bf
r},t)}\rangle,$
and $Q'({\bf r},t)$ the same quantity but in the primed system.

We now need to prove that Eq.~(\ref{difference}) admits only one
solution, i.e., $\Delta {\bf A}({\bf r},t)$ is completely determined
by the averaged dynamics of the current and particle densities, once
the coupling with the bath, $\hat V$, is assigned.
To this end we expand Eq.~(\ref{difference})
 in series about $t=0$ and obtain an equation for the
$l$-th derivative of the vector potential $\Delta {\bf A}({\bf
r},t)$. That one can expand this equation in a time series about
$t=0$ follows immediately from the analyticity of the vector
potential and Eq.~(\ref{operator-dynamics})~\footnote{If $\hat
V\equiv 0$, Eq.~(\ref{operator-dynamics}) reduces to the usual
Heisenberg equation of motion. The presence of the coupling $\hat V$
does not affect the analytic properties of the equations of motion
for the ensemble-averaged quantities.}. We thus arrive at the
equation
\begin{equation}
\begin{split}
\overline {n_0({\bf r})}(l+1)\Delta A_{l+1}({\bf r})=&-\sum_{k=0}^{l-1}(k+1)\overline {n_{l-k}({\bf r})}\Delta A_{k+1}({\bf r})\\
&+\Delta Q_l({\bf r})\\
&+\sum_{k=0}^l \overline {j_{l-k}({\bf r})}\times \left[\nabla\times
\Delta A_{k}({\bf r})\right] \label{expansion}
\end{split}
\end{equation}
where, given an arbitrary function of time $f({\bf r},t)$, we have
defined the series expansion
$f_l({\bf r})\equiv \frac{1}{l!}\left.\frac{\partial^l f({\bf
r},t)}{\partial t^l}\right|_{t=0}.$
We are now left to prove that the rhs of Eq.~(\ref{expansion}) does
not contain any term $\Delta A_{l+1}({\bf r})$. This follows from
the fact that the dynamics of any ensemble-averaged operator is
given by Eq.~(\ref{operator-dynamics}). Indeed, this implies that the $l$-th
time derivative of any operator can be expressed in terms of its
derivatives of order $k<l$, time derivatives of the Hamiltonian of
order $k<l$, and powers of the operators $\hat V$ and $\hat V^\dagger$. The time
derivatives of the Hamiltonian do contain time derivatives of the
vector potential ${\bf A}({\bf r},t)$, but always of order $k<l$.
Then on the rhs of Eq.~(\ref{expansion}) no time derivative of order
$l+1$ appears. Equation~(\ref{expansion}) can be thus viewed as a
recursive relation for the time derivatives of the vector potential
$\Delta {\bf A}({\bf r},t)$. To complete the recursion procedure we
only need to assign the initial value $\Delta {\bf A}_0({\bf
r})={\bf A}({\bf r},t=0) - {\bf A}'({\bf r},t=0)$. Since in the unprimed
and primed systems the densities and current densities are, by
hypothesis, equal, the initial condition is simply given by
$\overline {n({\bf r},t=0)}\Delta A_0({\bf r})=\langle
\Psi_0|\overline{\hat j_p (r,t=0)}|\Psi_0\rangle
-\langle
\Psi'_0|\overline{\hat j_p (r,t=0)}|\Psi'_0 \rangle$,
where $\hat j_p({\bf r})=(1/2m)\sum_i \{\hat p_i,\delta({\bf r}-\hat
r_i)\}$ is the paramagnetic current density operator.

The same considerations as in Ref.~\cite{Vignale2004} about the
finiteness of the convergence radius of the time
series~(\ref{expansion}) apply to our case as well. We rule out the
case of a vanishing convergence radius by observing that it seems
implausible that the smooth (in the ensemble-averaged sense)
dynamics induced by Eq.~(\ref{evolution}) can introduce a dramatic
explosion of the initial derivatives of $\Delta {\bf A}({\bf r})$.
If this holds, the expansion procedure~(\ref{expansion}) can be
iterated from the convergence radius time onward. We have then
proved that Eq.~(\ref{expansion}) completely determines the vector
potential $\Delta {\bf A}({\bf r},t)$ and thus, since ${\bf A}({\bf
r},t)$ is assumed known, it determines ${\bf A}'({\bf r},t)$
uniquely, up to a gauge transformation.

To finalize our proof, we consider the case in which $U=U'$ and
$|\Psi_0\rangle=|\Psi'_0\rangle$. If this holds, $\Delta {\bf
A}_0({\bf r})\equiv 0$. Then the recursion relation admits the
unique solution $\Delta {\bf A}_l({\bf r})\equiv 0$  for any $l$,
and at any instant of time $t$ we have ${\bf A({\bf r},t)}={\bf
A'({\bf r},t)}$ (still up to a gauge transformation).

{\it Discussion:} The theorem we proved provides solid grounds for
the development of a Kohn-Sham (KS) scheme for the study of the
dynamics of open quantum systems, i.e., many-particle systems coupled
to an external bath. Indeed, we can choose as the primed system a
non-interacting one, i.e., $U'=0$. We thus find that any
(ensemble-averaged) current density that is interacting {\bf
A}-representable is also non-interacting {\bf
A}-representable~\cite{Vignale2004,Giulianivignale}. (It is worth to
point out that, in general, the current density is not
$v$-representable, i.e., the mapping between current density and
external scalar potential is not invertible~\cite{DAgosta2005a}.)
The dynamics of the many-particle system can then be mapped into the
dynamics of an "auxiliary" KS Slater determinant evolving according to
\begin{equation}\label{evolutionKS}
\begin{split}
\partial_t \Psi^{KS}({\bf r},t)=&-i\hat H^{KS}(t)\Psi^{KS}({\bf r},t)-\frac{1}{2} \hat V^\dagger \hat V
\Psi^{KS}({\bf r},t)\\
&+\hat V\ell(t)\Psi^{KS}({\bf r},t),
\end{split}
\end{equation}
where
\begin{equation}
\hat H^{KS}(t)=\sum_i\frac{\left [ \hat p_i +e {\bf A}({\hat r_i},t)
+e {\bf A}_{xc}(\hat r_i,t)
\right ]^2}{2m}
+ V_H(\hat r_i,t),\label{hKS}
\end{equation}
with ${\bf A}_{xc}[\overline{ {\bf j}({\bf r},t)}, |\Psi_0\rangle,
\hat{V}]$ the xc vector potential, and $V_H({\bf
r},t)$ the Hartree potential. In actual calculations, one can apply,
as a starting point, available approximations for ${\bf
A}_{xc}$,~\cite{Vignale1996b,Vignale1997b, Kurzweil2005} but more
work in identifying the role of the bath on the correlations of the
system is necessary.

We conclude by noting that the present theorem shows that given a
bath and its interaction with the system, the ensemble-averaged expectation
value of any observable can be written in terms of the ensemble-averaged current
density. The most simple example is the average total current flowing
in the system. Indeed, given an arbitrary surface $S$ we can easily show that the
ensemble-averaged total current $\overline{I_{S}(t)}$ that flows
across that surface is
\begin{equation}
\overline{I_{S}(t)} =\int_S \overline{ {\bf j}({\bf
r},t)}\cdot d{\bf S} = \int_S \overline{ {\bf j}^{(KS)} ({\bf r},t)}\cdot
d{\bf S} =\overline{I^{(KS)}_{S}(t)}, \label{IKS}
\end{equation}
i.e., it is equal to the ensemble-averaged KS current
$\overline{I^{(KS)}_{S}(t)}$, where $\overline{ {\bf j}^{(KS)} ({\bf
r},t)}$ is the (ensemble-averaged) sum of expectation values of the
one-electron current-density operator in the populated KS states
$\psi_i^{KS}({\bf r},t)$. The result of Eq.~(\ref{IKS}) extends the
one proved in Ref.~\cite{DiVentra2004a} for finite and isolated
systems~\footnote{See also G. Stefanucci and C.-O. Almbladh, Phys.
Rev. B {\bf 69}, 195318 (2004).} to the case in which the system is
coupled to a bath. However, we note that the theorem we have proved
in this paper is not limited to any specific boundary condition on
the current density, and is thus valid for infinite systems as well.

In summary, we have introduced a TDCDFT theory for many-particle systems in
interaction with arbitrary external baths. We have named it
Stochastic TDCDFT. We have proved the uniqueness (up to a gauge
transformation) of the external vector potential, once the initial
condition and the bath operator are assigned. This theory greatly
expands the applicability of density-functional-based methods to
systems evolving beyond Hamiltonian dynamics. We thus expect it will
be of great value in the study of several physical problems, ranging
from optics to transport.

\acknowledgments
We thank K. Burke for useful discussions.
This work has been supported by the
Department of Energy grant DE-FG02-05ER46204.

\bibliography{mine,articles,books}
\end{document}